\documentstyle[aps,epsfig]{revtex}
\def\beqa{\begin{eqnarray}}
\def\eqar{\end{array}}
\def\beqar{\begin{array}}
\def\eqa{\end{eqnarray}}
\def\bars{\begin{eqnarray*}}
\def\ears{\end{eqnarray*}}

\def\beq{\begin{equation}}
\def\eq{\end{equation}}

\newcommand\lr[1]{{\left({#1}\right)}}

\newcommand{\bt}{\begin{tabular}}
\newcommand{\et}{\end{tabular}}
\newcommand{\bd}{\begin{displaymath}}
\newcommand{\ed}{\end{displaymath}\noindent}
\newcommand{\ec}{\end{center}}
\newcommand{\bc}{\begin{center}}

\newcommand{\g}{\gamma}

\begin{document}
\title{Hard jet probes in terms of colorless QCD dipoles}
\author{R. Peschanski\thanks{%
CEA, Service de Physique Theorique, CE-Saclay, F-91191 Gif-sur-Yvette Cedex,
France}}
\maketitle

\begin{abstract}
The forward jet playing the r\^ole of a hard probe in high energy QCD scattering 
is 
analyzed  in terms of QCD dipole (color-singlet $q \bar q$) configurations in 
the transverse
position space. There are  sizable differences  with  the analoguous $q \bar q$ 
configurations of a hard photon probe  which may lead to significant 
phenomenological 
consequences on the analysis of hard processes  using a forward jet  probe, e.g. 
at the 
Tevatron. A  geometrical interpretation of the resulting distribution in terms 
of black disk 
diffraction is proposed. 
\end{abstract}
\bigskip

{\bf 1.} Studies on deep-inelastic scattering experiments by virtual photon 
probes have 
shown the interest of using the formalism of $q \bar q$ wave-functions of 
the photon in the infinite momentum frame \cite {bj71}. Indeed, the theoretical 
predictions for small-x structure functions at the leading-log approximation and 
in the multicolor $1/N_c$ limit  can be expressed in terms of the probability of 
finding the $q \bar q$ configurations inside the photon probe  convoluted with 
the high-energy dipole-dipole cross-section \cite {ni94,mu94,na95}. This scheme 
provides  a useful description of the hard probe in the {\it configuration 
space} of the variables $\left(\vec r \equiv  r e^{i\phi}, z\right)$ where $r$ 
is the transverse size of the   $q \bar q$ state, $\varphi,$ the azimuthal 
orientation over the photon momentum and  $z$ (resp. $1-z$) is the energy 
fraction brought by the quark (resp. the antiquark) in the infinite momentum 
frame.

This scheme can be proved equivalent \cite {mu98} to the $k$-factorization 
property\cite {ca91}  of structure functions  calculated at small $x_{Bj}:$ 
a QCD photon vertex function is obtained convoluted with  the unintegrated 
off-shell gluon 
structure function solution of the Balitskii, Fadin, Kuraev, Lipatov (BFKL) 
equation \cite {bfkl}.  

For  high-energy QCD processes in general, the use of photon 
wave-functions are very useful for the investigations on the so-called 
``hard diffraction'' processes characterized by a large rapidity gap between 
particles produced at the photon and proton vertices at HERA. Indeed, the 
physical process by which a color singlet $q \bar q$ configuration of the photon 
propagates in the configuration space and interacts with the proton gives an 
interesting insight on the dynamics of ``hard diffraction'' initiated by a 
photon probe \cite {bi96}. 

In other reactions of interest, the hard probe is not furnished by a virtual 
photon. This is the case when an energetic forward jet with high transverse 
momentum $k_T$ serves as the hard QCD probe in various studies such as the 
Mueller-Navelet process \cite {mu86} and also  ``hard diffraction'' processes 
under study at the Tevatron by various combinations of jets and gaps (in the 
following we consider only one hard jet, independently of the way the transverse 
momentum is balanced, e.g.  by another jet). The 
forward jet as a hard probe is already present in  HERA studies where it allows 
a comparison with  perturbative QCD predictions for the hard-gluon on 
hard-photon scattering at high energy \cite {mu91}.  However, if a treatment of 
the forward jet or gluon vertex following the 
original BFKL scheme is known \cite{ba92}, the corresponding   formalism using
the propagation of $q \bar q$ color-singlets and their interaction with the 
target has not been yet discussed, to our knowledge. 
Our goal  is to derive an estimate  of the   color-singlet dipole content of a 
forward jet probe and to discuss its configuration space characteristics by 
comparison with the photon case. 

It is worthwhile to emphasize the similarities and differences expected between 
the singlet configurations of  photon and  gluon probes. In the photon case (see 
Fig.1a) the color singlet $q\bar q$ configurations are given by the light-cone 
probability distributions $\Phi^{\g}$ and interact {\it via} the dipole-dipole
cross-section $\sigma_d$ with the {\it primordial} dipole distribution in the 
target $\Phi^t.$ 

In the case of a forward gluon jet we want to consider, the 
situation is more delicate (see Fig.1b). The forward jet production is mediated 
by a virtual gluon (or a sea quark, which is equivalent in the multicolor 
$1/N_c$ 
limit of QCD we consider) which carries color and thus is not directly 
analyzable in terms of
color singlet $q\bar q$ configurations; It is more likely a three-body than
a two-body problem since it 
involves at least a soft exchanged gluon which carries away the color (see the 
Fig.1b). 
However, it is possible to make a reasonable 
assumption allowing to treat the problem in a way similar to the photon. Indeed, 
the BFKL 
interaction initiated by an initial (virtual) gluon can as well be described in 
terms of 
dipole-dipole interactions, e.g. in terms of color singlet configurations. 
Singularizing the most energetic dipole (which we identify as the {\it 
primordial} dipole in the forward jet formation), we can define its light-cone 
probability distribution $\Phi^{g},$ interacting through the same dipole-dipole 
cross-section $\sigma_d$ and the same vertex dipole distribution $\Phi^t$ as in 
the photon case. As we shall see, within the approximation that the 
abovementionned exchanged 
gluon carries little energy fraction to the remaining interacting dipoles, this 
QCD dipole factorization properties lead to a  prediction for $\Phi^{g},$ where 
the gluon is specified to be with transverse polarization only.
 
The plan of our study is the following. In the next section {\bf 2}, we briefly 
 introduce the appropriate QCD dipole formalism and derive the resulting 
prediction for the 
integrated Mellin transformed $\tilde \Phi^{g}$ of the  distribution $\Phi^{g}$ 
we look for.
Using a physically motivated  assumption on the overall energy carried by the 
{\it 
primordial} dipole, 
the last step of the derivation of $\Phi^{g}$ is made in section {\bf 3},  with 
a   
discussion on the properties of our result. Conclusion and questions 
for future work appear in section {\bf 4.}

\bigskip

{\bf 2.}  For a given target $t$, the transverse  projectile-target 
(where the projectile is either a photon or a gluon) total cross sections 
$\sigma^t$ obtained by high-energy QCD in the leading logs would read in the 
dipole scheme
\begin{equation}
\sigma^t=\int d^2r dz\;
\Phi(r,z;Q^2)\int d^2r_t dz_t\; 
\Phi^t(r_t,z_t;Q_t^2)\ \times \sigma_{d}(r,r_t;Y)\ ,
\label{definition}
\end{equation}
where we introduced the probability distributions $\Phi(r,z;Q^2)$ for 
the transversely  polarized projectile and  
$\Phi^t(r_t,z_t;Q_t^2)$ for the primordial  dipole distribution inside the 
target.
The dipole-dipole cross section $\sigma_d(r,r_t;Y)$ within the (large) rapidity 
range $Y$
can be expressed \cite {mu98} using an inverse Mellin transform:
\begin{equation}
\sigma_{d}(r,r_t;Y)=2\pi \alpha_s^2\int\frac{d\gamma}{2i\pi}
\frac {\left(r^2\right)^{1\!-\!\gamma}\left(r_t^2\right)^{\gamma}}
{\g^2 (1\!-\!\gamma)^2}\ 
e^{\epsilon(\gamma)Y}\ ,
\label{dipoles}
\end{equation}
where 
\begin{equation}
\epsilon(\gamma)=\frac{\alpha_s N_c}{\pi}\ 
\left[2\Psi(1)-\Psi(\gamma)-\Psi(1\!\!-\!\!\gamma)\right]\ 
\label{eqn:kernel}
\end{equation}
 is the (Mellin transformed) BFKL kernel \cite{bfkl} with $\Psi(\g)\equiv 
d\log\Gamma/d\g.$
 
Let us come to our derivation. 
Since in formula 
(\ref{definition}) the 
target distribution $\Phi^t(r_t,z_t;Q_t^2)$ can be expanded over  individual 
massive dipole components  $\Phi^t\equiv1/\pi\ \delta(r^2_t-1/Q^2_t) 
\delta(z_t-1/2),$  we shall restrict ourselves to the gluon-dipole cross-section 
writing
\begin{equation}
\sigma=
  \frac{4\pi \alpha^2_s}{Q^2}\int\frac{d\gamma}{2i\pi}
  \left(\frac Q{Q_t}\right)^{2\gamma}\!\! 
  e^{\epsilon(\gamma)Y}\frac {\tilde \phi^g(\gamma)}{\g^2 (1\!-\!\gamma)^2}
 \ ,
\label{crossssection}
\end{equation}
where
\begin{equation}
\int d^2r \left(r^2Q^2\right)^{1\!-\!\gamma}\int dz
\>\Phi^g(r,z;Q^2)=\tilde \phi^g(\gamma)
\label{eqn:defphi}
\end{equation}
is the integrated (over the  energy fraction $z$) and mellin-transformed (over 
the transverse 
coordinate space) 
of the {\it 
primordial} dipole distribution out of an incident virtual gluon.

The BFKL solution for the same gluon-dipole cross-section writes~\cite 
{na95,mu98}
\begin{equation}
\sigma=
  \frac{\pi^2 \alpha^2_s}{2Q^2}\int\frac{d\gamma}{2i\pi}
  \left(\frac Q{Q_t}\right)^{2\gamma}\! 
  e^{\epsilon(\gamma)Y} \ \frac {v(\g)}{(1\!-\!\g)}\ .
\label{crossbfkl}
\end{equation}
It is obtained using the {\it unintegrated}
 $k$-factorized gluon coupling to a $q\bar q$ pair which is 
given{\footnote{These {\it unintegrated}
couplings can be deduced from the factorization of the elementary 
dipole-dipole cross-section (\ref{dipoles}) since 
$v(\gamma)v(1\!-\!\g)\equiv 1/16\g^2 (1\!-\!\gamma)^2,$ where $v(\g)$ 
corresponding to 
formula (\ref{gcoupling}) is for the lower vertex and $v(1\!-\!\g)$ is 
for the upper one.} (for the lower vertex)
by
\begin{equation}
v(\gamma)=2^{-\!2\gamma\!-\!1}\ \frac{\Gamma(1\!-\!\gamma)}{\gamma 
\Gamma(1\!+\!\gamma)}\ .
\label{gcoupling}
\end{equation}
Note the factor $(1\!-\!\g)^{-1}$ in the integral (\ref{crossbfkl}) due to  the 
expression of 
cross-sections usually given after integration over $k_T$ above a certain value 
$Q$ which 
plays the r\^ole of the virtuality~(see, e.g. \cite {mu98}). 

It is 
important to notice a subtlety about the incoming gluon which  initiates the 
process. The 
incident gluon acquires a different status than 
 the exchanged ones appearing in the BFKL derivation of $\sigma_d.$ Indeed, it 
is known 
that the BFKL gluons are reggeized and predominantly longitudinally polarized in 
the axial gauge \cite {mu94}. On the other hand the incident gluon is  
predominantly transversely polarized in the same gauge \cite {do91} since it 
comes from the DGLAP evolution from the physical projectile (e.g. a proton). 
This non trivial helicity-flip is not contradictory to the properties of the  
BFKL Pomeron 
(which contains the DGLAP evolution at the double leading log accuracy) since it 
can be seen from the high-energy properties
of the gluon propagators and polarization vectors in the BFKL derivation~\cite 
{wa96}. From 
that 
property comes the fact that we single out our prediction for the 
transversely polarized gluon only.

By consistency of both equivalent 
formulations (\ref{crossssection},\ref{crossbfkl}), in the same way as it has 
been done for 
the photon case~\cite{mu98}, one  gets  the following 
result
\begin{equation}
\tilde \phi^g(\gamma) = 8\g^2(1\!-\!\gamma)\  v(\gamma)\equiv  2^{2\!-\!2\g}
\frac {\Gamma(2\!-\!\g)}{\Gamma(\g)}\ .
\label{relationg1}
\end{equation}
This expression is to be compared with the analogous one for the photon 
namely~\cite{na95,mu98}:
\begin{equation}
\tilde \phi^{\g} = C \frac 1 {\g(1\!-\!\g)}\ \frac 
{\Gamma^3(1\!+\!\g)\Gamma(2\!+\!\g)\Gamma(2\!-\!\g)\Gamma(3\!-\!\g)}{\Gamma(2\!+
\!2\g)\Gamma(
4\!-\!2\g)}\ ,
\label{relationph1}
\end{equation}
where the 
normalization factor ($C= \alpha_{em}N_c e_q^2 /2\pi^2$ with $e_q$ the quark 
charge) takes into account in this case the color, flavor and 
electromagnetic charges.

\bigskip

{\bf 3.} The solution for the photon case, corresponding to the inversion of the 
integrated 
and Mellin transformed 
function $\tilde \phi^{\g}$ of formula (\ref{relationph1}), is well 
known~\cite{bj71,ni94}. 
The probability distribution of the light-cone  color-singlet $q \bar q$ 
configurations for a 
transversely 
polarized virtual photon   writes
\begin{equation}
\Phi^{\g} \lr{z,r,Q^2}  \equiv C \left( z^2+\lr{1\!-\! z}^2\right)\ \hat Q^2 \  
\tilde\phi^{\g} \lr{u} \ ;\  \tilde\phi^{\g} \lr{u} =
K^2_{1} \lr{u}
,
\label{phormulation}
\end{equation}
where $K_{1}$ is the  Bessel function of second kind and  by definition 
$u\equiv r\hat Q  \equiv rQ\sqrt{z(1\!-\!z)}.$ The $z$-dependent prefactors are 
known~\cite{bj71} to  come from the coupling of 
the quark and antiquark spinors to the transverse helicities of the virtual 
photon. Our task is now to find a similar inversion procedure for  formula 
(\ref{eqn:defphi}), i.e. applicable to the gluon 
distribution.

To our knowledge, the $z$-dependence of the $q\bar q$ system of the {\it  
primordial} dipole 
from a gluon probe is not known. We shall make use in the subsequent derivation 
of a 
plausible assumption (at least for $z$ not too near $(0,1),$ namely that the 
{\it primordial} 
dipole in the gluon jet retains the most part of the energy and longitudinal 
momentum 
of the initial gluon. This assumption can be based  on the physical idea that 
the color 
degrees of freedom are much more ``volatile'' than the energy 
momentum\footnote{The same 
hypothesis has led to interesting models for ``hard diffraction'' connecting  
color-singlet 
production to the  kinematics of perturbative vertices \cite {bu91}.}, which 
means that the
energy balance in a local interaction $g\to (colored)q\bar q$ is not changed 
when transmuted 
to the system $g\to (soft)g+(singlet)q\bar q$ as schematized in Fig.1b. In other 
terms we 
assume a
hierarchical structure of 
the cascade of interacting dipoles in the rapidity space which keeps the most 
part of energy 
in the first rank in rapidity, i.e. the {\it primordial} dipole. 
In practical terms, the exchanged 
colored gluon (see the upper vertex in Fig.1b) brings a negligeable energy 
fraction out of the quark (or antiquark) to which it is coupled and thus the  
balance in the infinite momentum 
frame between the energy 
fraction $z$ brought by the quark and $1\!-\!z$ for the antiquark remains 
approximatively the same as for the direct coupling of the gluon to $q\bar q.$

Thus, in this framework, we are led to  the same kinematical prefactors 
for the gluon as for the photon  
(\ref{phormulation}), namely we write 
\beq
\Phi^{g}(r,z;Q^2) = \left( z^2+\lr{1\!-\! z}^2\right)\ 
\hat Q^2\ \phi^{g} \lr{u}
\label{form}
\eq
With this ansatz, the solution boils down to solve an 
 inverse Mellin transform which gives:
\beqa
\phi^{g} \lr{u}  &=& \frac 1{\pi}\int\frac{d\gamma}{2i\pi} u^{2(\g\!-\!2)}
\ 2^{1\!-\!2\g}\frac 
{\Gamma(2\!-\!\g)\Gamma(2\g\!+\!2)}{\Gamma(\g\!+\!2)\Gamma^2(\g)}
=  \frac 5{2\pi}\ _2F_3\left(\frac 72,3;4,2,2,-u^2\right)
\nonumber\\
&\equiv& \frac 1{\pi}\left\{ \frac {J_1^2 \lr{u}}{u^2}+J_0^2 \lr{u}-J_1^2 
\lr{u}-\frac 
{J_2^2 
\lr{u}}{2u^2}\right\}\ ,
\label{gluonfonction}
\eqa
where  $J_{0,1,2}$ are the usual Bessel functions of first kind. The 
normalization is fixed in such a way that $\int d^2r dz
\>\Phi^g(r,z;Q^2)\equiv 1.$ the hypergeometric function $_2F_3$ has thus been 
finally 
cast~\cite{prudnikov} 
into a  form similar to the photon case, but with a combination of Bessel 
functions of first 
kind instead of the squared of the Bessel function of second kind appearing for 
the photon in
(\ref{phormulation}). 

  In the following discussion
we shall comment on the interesting differences with the photon case 
(\ref{phormulation}) which appear clearly on Fig.2, where both functions $\tilde 
\phi$ have 
been drawn. 

{\bf i)} Fig.2 shows the function $\phi^{g}$ compared to $\phi^{\g}$ as a 
function of $u^2=Q^2z(1\!-\!z)$ in the region $u^2\le 2.$ The existence of 
regions with negative  $\phi^{g}$ beyond this $u^2 = 2$  
shows that a simple interpretation in terms of probability distributions -as 
for the photon case~\cite{bj71}- cannot be valid. We shall dicuss later a 
possible 
interpretation of this fact.

{\bf ii)} Fig.2 also reveals that the space-time configuration of the {\it 
primordial}
singlet state resulting from the virtual gluon is likely to be much more spread  
in 
configuration space
than in the photon case. Indeed, the combination of Bessel functions in formula 
(\ref {gluonfonction}) has a  tail behaving as $\cos (2u)/u,$ while the photon 
distribution tail is exponentially small as $K_1^2(u)\sim\exp (-2u)/u.$ The 
difference is already visible at finite values of $u62$  as shown by Fig.2. As a 
consequence, for a given scale $\hat Q$ of  the hard probe, there can be
larger fluctuations in the dipole size than for the photon. Thus, even a 
reasonably hard scale probe can give rise to large dipoles for which, e.g., 
saturation 
and/or screening corrections may be more important than for the photon. This 
remark may be phenomenologically relevant when comparing
hard QCD processes at high energy at the Tevatron  and at Hera. 

{\bf iii)}
The formula (\ref {gluonfonction}) with both positive and negative contributions 
may have 
an interesting interpretation in terms of real and virtual 
contributions of simple geometrical meaning. Indeed, it is well known that 
the dipole wave-function formalism~\cite{mu94} implies both terms. It is thus 
conceivable that factorizing out the contribution of the {\it primordial} dipole
keeps this feature. Note that  the virtual contributions are necessary to
obtain an overall equivalence between the dipole model and the BFKL formalism, 
even if, separately, real 
and virtual contributions are different in both approaches~\cite{mu96}.

{\bf iv)} A nice geometrical interpretation of our resulting formula 
(\ref {gluonfonction}) comes from the comparison with the well-known 
 amplitudes for diffraction by a black 
disk~\cite{austern}. In a high-energy version taking into account also the 
helicity 
structure~\cite{reggeometry} the black disk amplitudes write
  \begin{equation}
M^{el}_n(\hat Q)= \hat Q^2\int_0^{\infty}\ b^{n}db^2 J_n(b\hat Q)\ 
\theta(b^2\!-\!r^2) \propto  \frac {J_{n+1}(r\hat Q)}{r\hat Q}\ ,
\label{elastic}
\end{equation}
for elastic amplitudes on the bulk of the black disk and 
  \begin{equation}
M^{inel}_n(\hat Q)= \hat Q^2\int_0^{\infty}\ b^{n}db^2 J_n(b\hat Q)\ 
\delta(b^2\!-\!r^2) \propto   {J_{n}(r\hat Q)}\ ,
\label{inelastic}
\end{equation}
for inelastic amplitudes corresponding to peripheral reactions on the edge of 
the disk. $n$ is the net helicity flip in the 
reaction~\cite{reggeometry}. 

If we are allowed to interpret the Bessel functions appearing in 
(\ref {gluonfonction}) as
  wave-function components (both for real and virtual contributions), a 
geometrical interpretation would be  that the {\it primordial} dipole totally 
{\it absorbs} the partial waves for  impact 
parameter $b\le r,$ where $r$ is its size. 
\bigskip

{\bf 4.} In conclusion,
assuming the quasi-two-body kinematical coupling of an initial virtual gluon to 
the {\it 
primordial dipole} of cascading dipoles describing the BFKL amplitudes, we have 
derived the analogue of the transverse photon probability distribution 
(\ref{phormulation}) for a transverse gluon, see formulae 
(\ref {gluonfonction}) . The solution differs from the photon, in 
particular
by its wider extension in configuration space, allowing for larger fluctuations 
of the dipole sizes for a given gluon virtuality.  An interpretation 
in terms of black disk absorption has been proposed.

Some questions arise from our analysis.

-Can we discuss the existence, properties and limitations of a configuration 
space distribution $\phi^g$ {\it directly} from the three-body 
problem~\cite{mu20} 
$g \to (soft)g+ singlet (q \bar q ),$ corresponding to the coupling of the 
initial gluon to 
the {\it primordial} dipole and another soft gluon bringing out the color 
quantum numbers, 
see Fig.1?

-Are the real/virtual states decomposition suggested by our resulting formula 
valid and does it corresponds (in the QCD dipole basis) to the existence of 
{\it primordial}  colored states in the gluon wave-function?
\eject
-Can we say something about the longitudinal polarization of the gluon by 
comparison with the known result for the photon\footnote{We remark that assuming 
by example 
the same dipole coupling~(\ref{crossssection}) for both polarizations of the 
gluon leads also 
to a simple decomposition in terms of bessel functions, namely
$$
\Phi_{L}^{g} \lr{z,r,Q^2}   \equiv  \frac 3 {\pi} z\lr{1\!-\! z}\ \hat Q^2 
\ \left\{\frac {J_1^2 \lr{u}}{u^2}\!+\!J_0^2 \lr{u}\!-\!\frac 23J_1^2 \lr{u} 
\!-\!\frac 
13J_2^2 \lr{u}\right\}\ .
$$
} ?

We think worthwhile to address these questions, since the phenomenology of
forward jets as QCD hard probes, useful e.g. at the Tevatron, requires a 
better understanding of the virtual gluon in configuration space. This could 
shed a new light 
(or give new tools) to high energy and diffractive QCD scattering at hadron 
colliders and 
perhaps help understanding the physical origin of  non-factorization or 
saturation 
mechanisms for jet-induced hard QCD scattering.
 
 \bigskip
\bigskip
{\bf Acknowledgements}

\noindent We acknowledge fruitful discussions with Henri Navelet, St\'ephane 
Munier and Christophe Royon.

\eject
{\bf FIGURES}

\bigskip  

\begin{center}
\psfig{figure=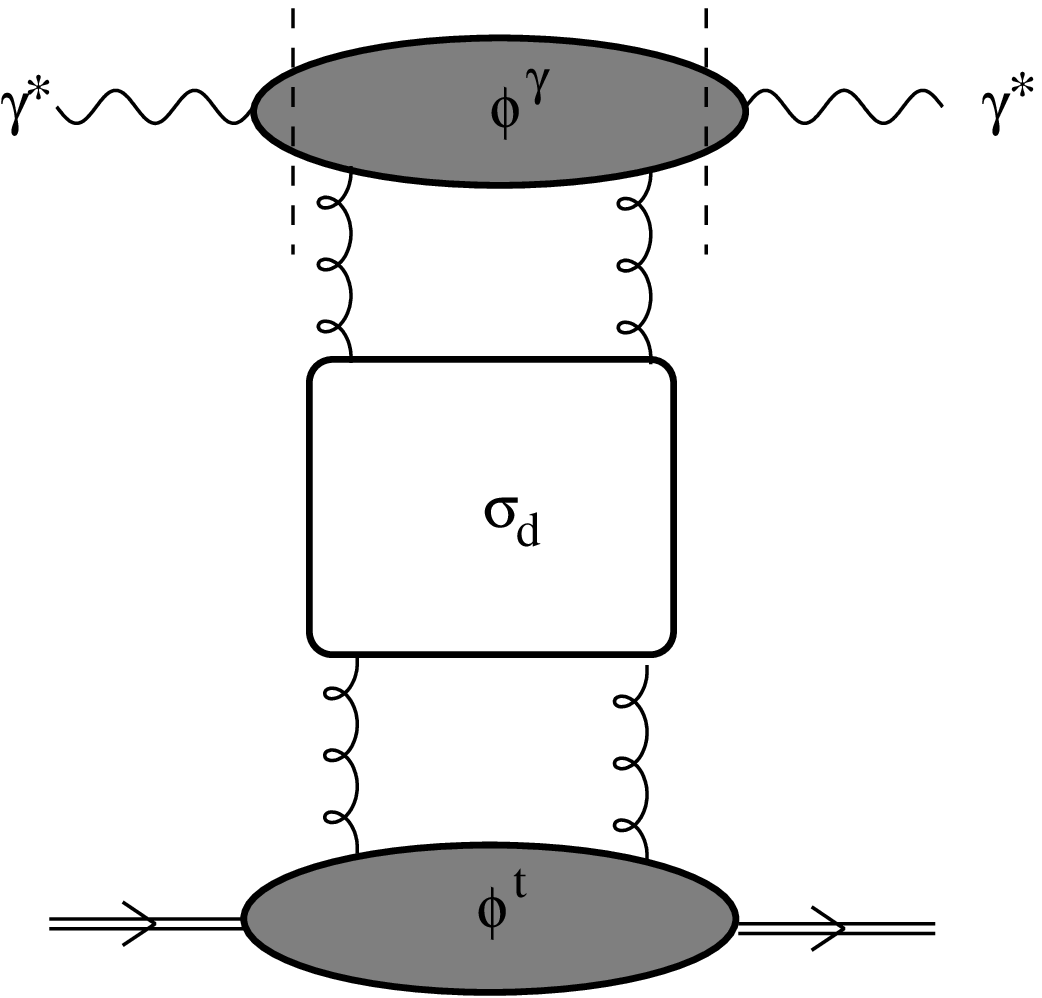,width=7cm}\\
\vskip 0.5cm
{\large\bf a.}\\
\vskip 1cm
\psfig{figure=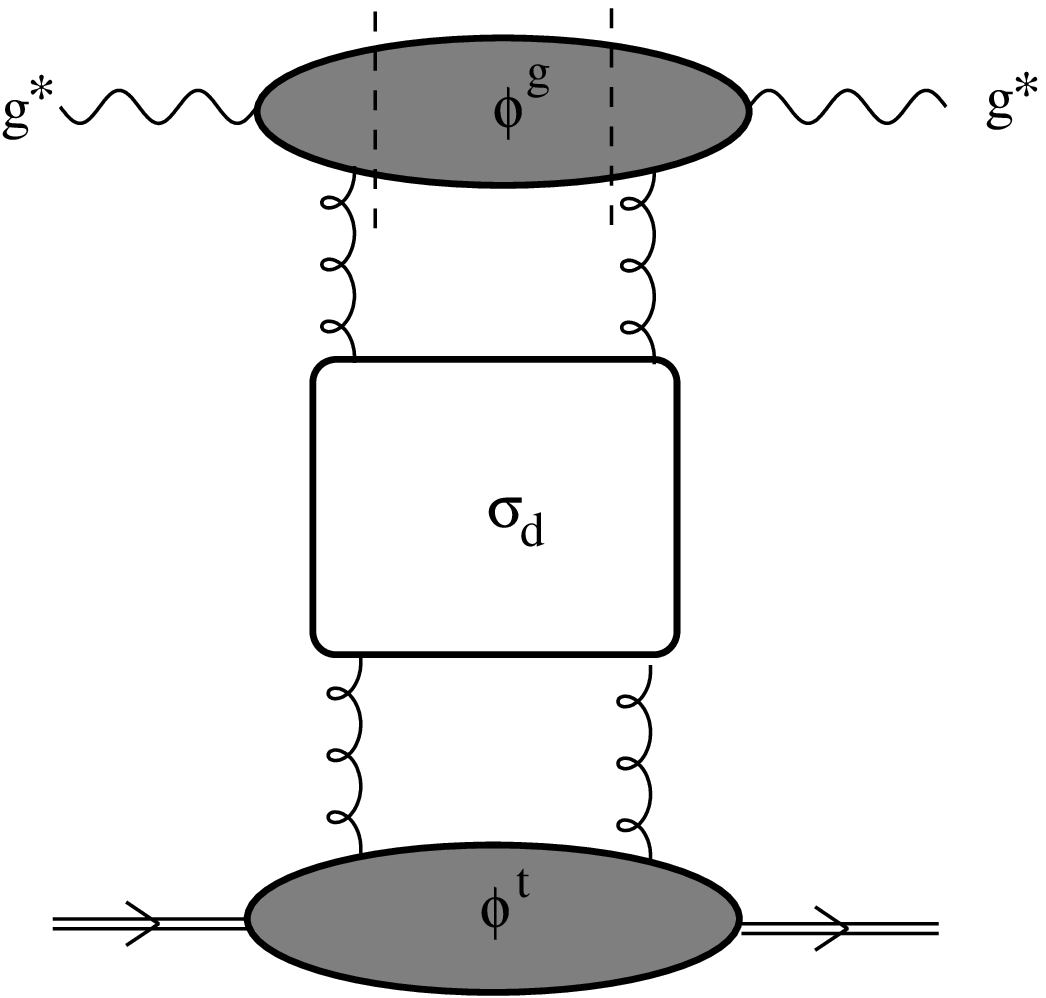,width=7cm}\\
\vskip 0.5cm
{\large\bf b.}\\
\vskip 0.5cm
{\large\bf Figure 1}\end{center}

{\it Photon and gluon-initiated reactions in the QCD dipole basis}

The figures schematically describe the QCD dipole contributions to photon and 
gluon-induced reactions. 
\noindent Fig 1a: the photon-induced reaction; Fig 1b: the gluon-induced 
reaction. The photon (resp. gluon) probability distribution is denoted by 
$\phi^{\g}$ (resp. $\phi^g$). $\phi^{t}$ is the target dipole distribution 
and $\sigma_d$  the dipole-dipole cross-section, see the corresponding formulae 
in the text.

In both cases the intermediate singlet $q \bar q$ state discussed in the text
is marked by a dotted line.

\bigskip  
\eject
\begin{center}
\psfig{figure=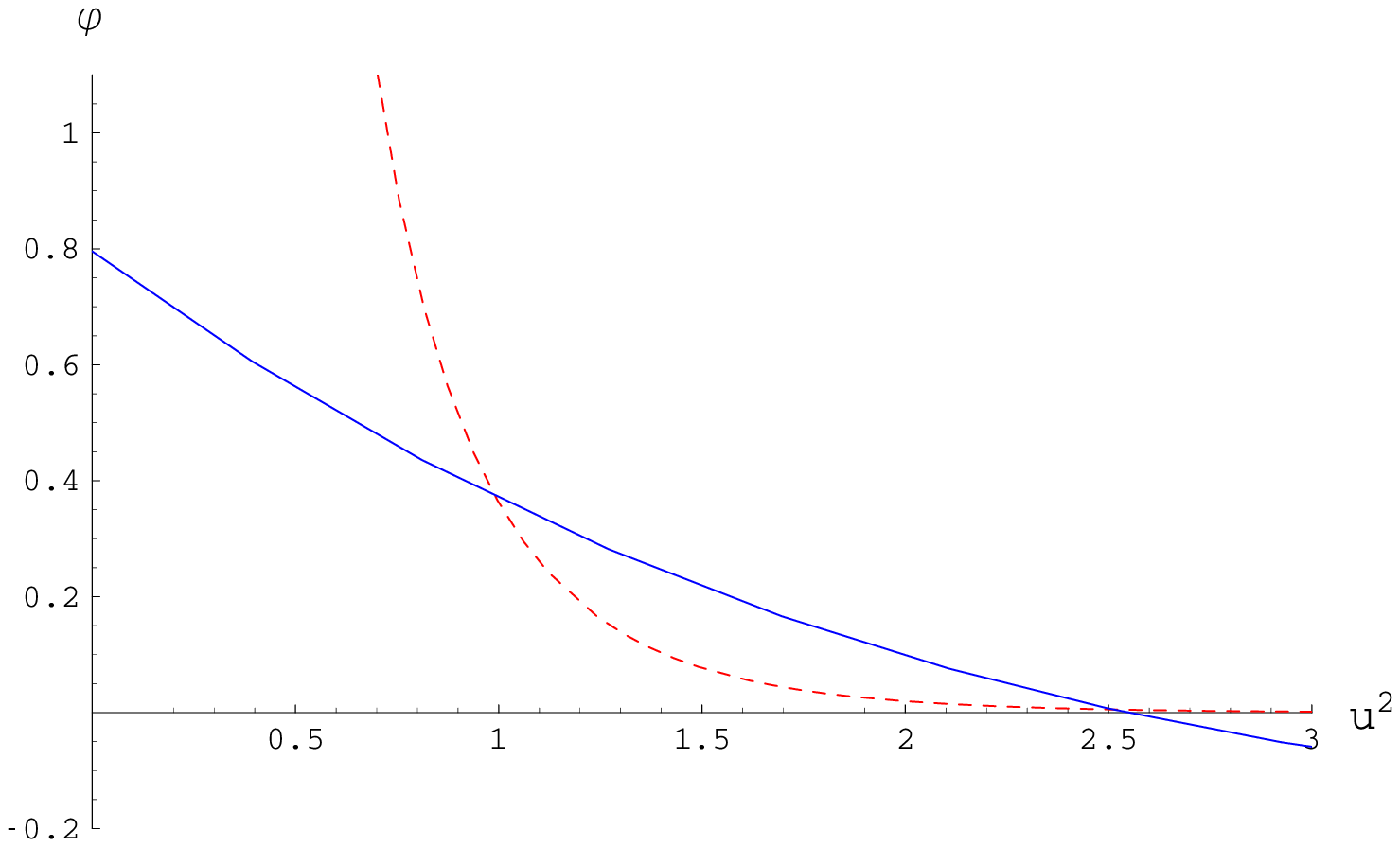,width=14cm}\\

\vskip 0.5cm
{\large\bf Figure 2}\end{center}

{\it The gluon probe vs. the photon probe}

The figure represents the gluon transverse polarization distribution $\phi^g,$ 
formula 
(\ref{gluonfonction}), compared to $\phi^{\g},$ formula (\ref{phormulation}), 
for the photon 
(with the constant C equal to 1 for making easier the comparison). The abcissa 
is defined by 
$u^2\equiv r^2Q^2z\lr{1-z}.$ 

\noindent Continuous line: the gluon probe distribution; Dashed line: the 
photon probe distributions.

\begin{references}

\bibitem{bj71}
J.D.Bjorken, J.Kogut and Soper, {\it Phys.Rev.} {\bf D3} (1971) 1382.
.
\bibitem{ni94}
N.N.Nikolaev, B.G.Zakharov, {\it Zeit. f\"ur. Phys.} {\bf C49} (1991) 607;
{\it Phys. Lett.} {\bf B332} (1994) 184.

\bibitem{mu94}
A.H.Mueller, {\it Nucl. Phys.} {\bf B415} (1994) 373;
A.H.Mueller and B.Patel, {\it Nucl. Phys.} {\bf B425} (1994)
471;
A.H.Mueller, {\it Nucl. Phys.} {\bf B437} (1995) 107.

\bibitem{na95}
H. Navelet, R. Peschanski, Ch. Royon, {\it Phys. Lett.} {\bf B366} (1995) 329.
H. Navelet, R. Peschanski, Ch. Royon, S. Wallon, {\it Phys. Lett.} 
{\bf B385} (1996) 357.

\bibitem{mu98}
S. Munier and R. Peschanski, {\it Nucl. Phys.} {\bf B524 }(1998) 377.

\bibitem{ca91}
S. Catani, M. Ciafaloni, F. Hautmann, {\it Nucl. Phys}. {\bf B366} (1991) 135.
J. C. Collins, R. K. Ellis, {\it Nucl. Phys.} {\bf B360} (1991) 3. E. M. Levin, 
M. G. Ryskin, Yu. M. Shabelskii, A. G. Shuvaev,
{\it Sov. J. Nucl. Phys.} {\bf 53} (1991) 657.


\bibitem{bfkl}  L.N.Lipatov, {\it Sov. J. Nucl. Phys.} {\bf 23} (1976) 642;
V.S.Fadin, E.A.Kuraev and L.N.Lipatov, {\it Phys. lett.} {\bf B60} (1975)
50; E.A.Kuraev, L.N.Lipatov and V.S.Fadin, {\it Sov.Phys.JETP} {\bf 44}
(1976) 45, {\bf 45} (1977) 199; I.I.Balitsky and L.N.Lipatov, {\it %
Sov.J.Nucl.Phys.} {\bf 28} (1978) 822.

\bibitem{bi96}  A.Bialas, R.Peschanski, {\it Phys. lett.} {\bf B378} 302; {\bf 
B387} 405 (1996); For a recent review, see e.g. A.Hebecker {\it Diffraction in 
Deep Inelastic Scattering} hep-ph/9905226.


\bibitem{mu86} A.H.Mueller and H.Navelet,  {\it Nucl. Phys.} {\bf B282} (1987) 
107.


\bibitem{mu91}  A.H.Mueller, {\it Nucl. Phys.} {\bf B} (Proc. Suppl.) 18C (1991) 
125.

\bibitem{ba92}
J.Bartels, A.De Roeck, M.Loewe, {\it Zeit. f\"ur Phys.} {\bf C54} (1992) 921; 
W-K. Tang, {\it Phys. lett.} {\bf B278} (1992) 635; 
J.Kwiecinski, A.D.Martin, P.J.Sutton, {\it Phys.Rev.} {\bf D46} (1992) 921.
 
\bibitem{do91}
See, for instance, Yu.L.Dokshitzer, V.A.Khoze, A.H.Mueller, S.I.Troyan 
{\it Basics of perturbatice QCD} (Editions Fronti\`eres, J.Tran Thanh Van Ed. 
1991).
 
\bibitem{wa96}
See, for instance, S.Wallon, PHD Thesis, Orsay University, France (in French).

 
\bibitem{bu91}
W.Buchmuller, A.Hebecker, {\it Phys. lett.} {\bf B355} (1995) 573. A.Edin, 
G.Ingelman, 
J.Rathsman, {\it Phys. lett.} {\bf B366} (1996) 371; {\it Z. Phys.} {\bf C75} 
(1997) 57; {\it 
Phys. Rev.}  {\bf D56} (1997) 7317; see A.Hebecker in 
ref.~\cite{bi96} for a review and complete list of references;


\bibitem{prudnikov}
A. Prudnikov, Y. Brychkov and O. Marichev,
{\it Integrals and Series} (Gordon and Breach Science Publishers, 1986).


\bibitem{mu96}  Cheng and A.H.Mueller, {\it Nucl. Phys.} {\bf B415} (1994) 373.

\bibitem{austern} N.Austern {\it Direct nuclear reactions theories}  
Interscience monographs and texts in physics and astronomy, vol. XXV 
(R.E.Marshak, ed., Wiley-interscience 1970).

\bibitem{reggeometry}  J.P.Ader, G.Cohen-Tannoudji, C.Gilain, R.Lacaze and 
R.Peschanski, {\it Il Nuovo Cimento} {\bf 27A} (1975) 385.


\bibitem{mu20}
For an interesting approach see S.Munier, to appear.
\end{references}
\end{document}